\documentclass[nofootinbib,twocolumn,showpacs,preprintnumbers,pre,aps,superscriptaddress]{revtex4-1}

\usepackage[dvipdfmx]{graphicx}
\usepackage{bm}
\usepackage{amsmath}
\usepackage{amssymb}
\usepackage{color}

\begin{document}

\title{Onsager's Variational Principle for Nonreciprocal Systems with Odd Elasticity}

\author{Li-Shing Lin}
\affiliation{
Department of Chemistry, Graduate School of Science,
Tokyo Metropolitan University, Tokyo 192-0397, Japan}

\author{Kento Yasuda}
\affiliation{
Research Institute for Mathematical Sciences, 
Kyoto University, Kyoto 606-8502, Japan}

\author{Kenta Ishimoto}

\affiliation{
Research Institute for Mathematical Sciences, 
Kyoto University, Kyoto 606-8502, Japan}

\author{Yuto Hosaka}

\affiliation{
Max Planck Institute for Dynamics and Self-Organization (MPI DS), 
Am Fassberg 17, 37077 G\"{o}ttingen, Germany}

\author{Shigeyuki Komura}\email{komura@wiucas.ac.cn}

\affiliation{
Wenzhou Institute, University of Chinese Academy of Sciences, 
Wenzhou, Zhejiang 325001, China} 

\affiliation{
Oujiang Laboratory, Wenzhou, Zhejiang 325000, China}

\affiliation{
Department of Chemistry, Graduate School of Science,
Tokyo Metropolitan University, Tokyo 192-0397, Japan}


\begin{abstract}
Using Onsager's variational principle, we derive dynamical equations for a nonequilibrium active system   
with odd elasticity. 
The elimination of the extra variable that is coupled to the nonequilibrium driving force leads to 
the nonreciprocal set of equations for the material coordinates. 
The obtained nonreciprocal equations manifest the physical origin of the odd elastic moduli 
that are proportional to the nonequilibrium force and the friction coefficients.
Our approach offers a systematic and consistent way to derive nonreciprocal equations for active 
matter in which the time-reversal symmetry is broken. 
\end{abstract}

\maketitle

Active matter is composed of self-driven units that consume energy from the environment and drive the 
system into out-of-equilibrium situations~\cite{Marchetti2013HydrodynamicsMatter,Hosaka22}. 
The motion of active particles is typically directional because the activity and local orientation lead to  
the broken time-reversal symmetry~\cite{Vicsek1995PHParticles}.
A practical way to describe the collective behaviors of active particles is to write down equations
for slow variables by considering all possible hydrodynamic force and flux 
pairs~\cite{Toner2005HydrodynamicsFlocks,Callan-Jones2011HydrodynamicsGels,Lin2021DynamicsAggregate}.

Concerning the dynamics of passive systems, it is known that Onsager's variational principle 
(OVP) provides us with proper dynamical equations that automatically satisfy Onsager's reciprocal relations 
and the second law of thermodynamics~\cite{Onsager1930MBAUAAYI}.
In this formulation, dynamical equations are obtained by minimizing a Rayleighian that is the sum of 
dissipation function and time derivative of free energy~\cite{Doi2011OnsagersMatter,DoiSoftMatterPhysics}.
OVP has been proven powerful for various passive dynamical problems in soft matter such as polymer 
systems~\cite{Doi21}, membranes~\cite{Okamoto16}, and liquid droplets~\cite{Man17}.
Nevertheless, OVP has not yet been employed for active systems due to the difficulties in determining the free energy 
or the dissipation function under nonequilibrium conditions.
One attempt is to introduce an additional term in Rayleighian accounting for the work power done by the active 
forces~\cite{Wang2021OnsagersMatter,Wang2022VariationalSolids}.

In this Letter, we focus on the new concept of active systems, i.e., odd elasticity introduced by Scheibner 
\textit{et al.}~\cite{Scheibner2020OddElasticity,Fruchart22}. 
Odd elasticity arises from an anti-symmetric (odd) component of the elastic modulus tensor.
It violates the energy conservation law and thus can exist only in active materials. 
Unlike passive materials, a finite amount of work can be extracted in odd elastic systems through quasi-static 
cycle of deformations~\cite{Scheibner2020OddElasticity,Fruchart22}.
Recently, the present authors proposed a thermally driven microswimmer with odd elasticity and  
demonstrated that it can exhibit a directional locomotion~\cite{YHSK21,Ishimoto22}.
We also showed that anti-symmetric parts of the time-correlation functions in odd Langevin systems are 
proportional to the odd elasticity~\cite{YIKLSHK22}.
Moreover, the concept of odd elasticity can be extended to quantify the nonreciprocality of active micromachines 
such as enzymes and motor proteins~\cite{Yasuda2022TheElasticity,Kobayashi22}.

The purpose of our work is twofold. 
First, we show that typical dynamical equations of an active system can be obtained within the framework
of OVP, which has a different structure compared with the previous 
approach~\cite{Wang2021OnsagersMatter,Wang2022VariationalSolids}. 
In addition to the regular material coordinates, we introduce an extra coordinate that is conjugate to the 
nonequilibrium active force (or torque) in the free energy. 
Being constantly driven by the active force, this extra coordinate transmits energy to the material coordinates 
through dissipative processes. 
Importantly, we obtain nonreciprocal (or nonconservative~\cite{Fruchart2021Non-reciprocalTransitions}) 
dynamical equations for the material coordinates after eliminating the extra coordinate from the coupled 
equations derived from OVP.

Second, the obtained nonreciprocal (coarse-grained) equations for the material coordinates 
provide us with the physical origin of the odd elasticity that was previously introduced merely by  
symmetry consideration~\cite{Scheibner2020OddElasticity,Fruchart22}.
We explicitly show that the odd elastic moduli are proportional to the nonequilibrium active force
and the friction coefficients appearing in the dissipation function of OVP.
We demonstrate our argument first by using a simplified model with only two material degrees 
of freedom and then later generalize it to a continuum model. 
Although we limit our discussion only to odd elastic systems, our approach of using OVP and nonequilibrium 
extra variable offers a systematic way to obtain nonreciprocal equations for other active systems.

\begin{figure}[tb]
\centering
\includegraphics[scale=0.2]{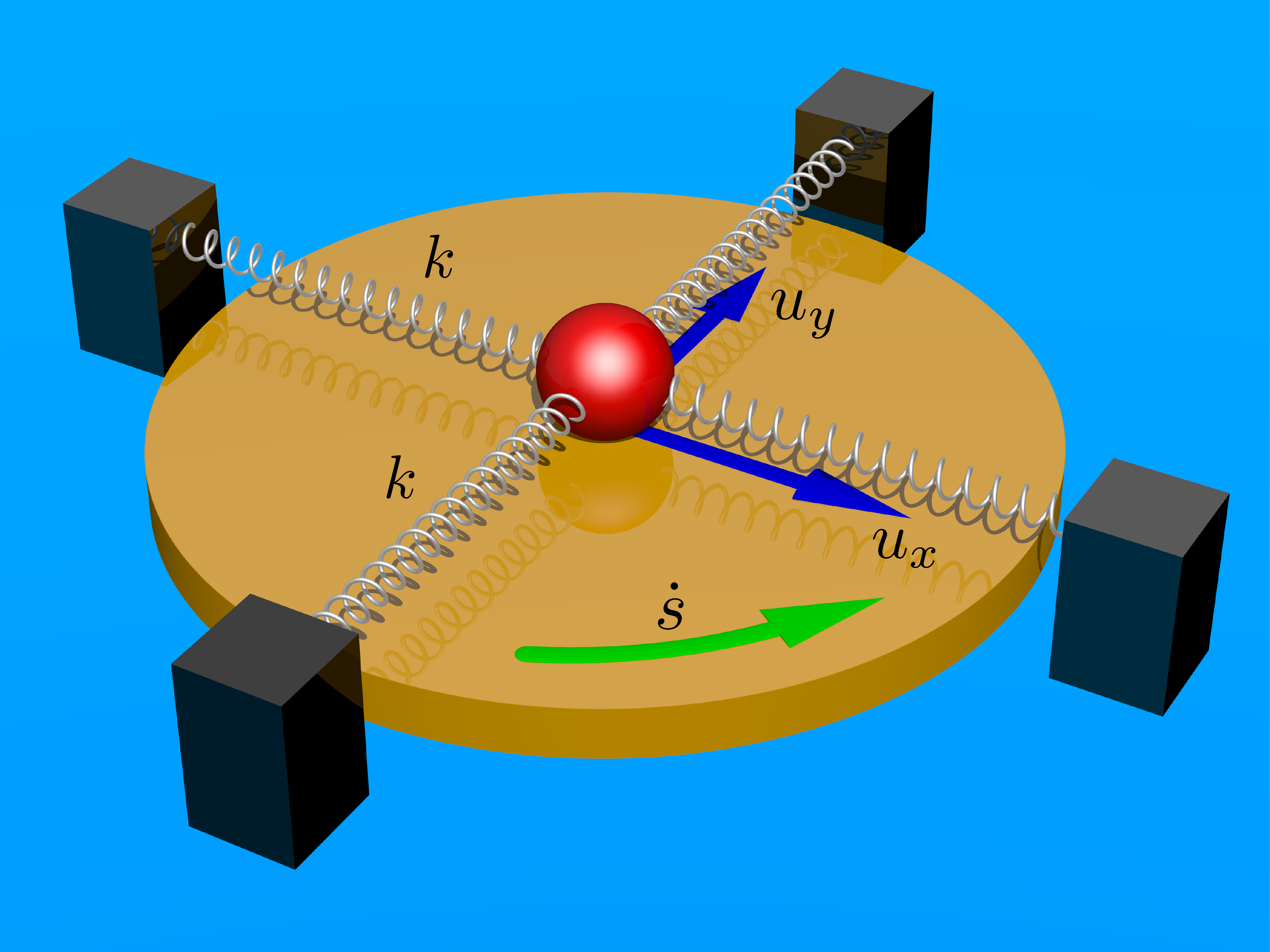}
\caption{(Color online) 
A particle confined by the four springs of stiffness $k$ and moving above a disk rotating with an 
angular velocity $\dot{s}$.
The displacements of the particle are denoted by $u_x$ and $u_y$.
The whole setup is immersed in a viscous fluid that causes dissipation.}
\label{fig_eg}
\end{figure}

To demonstrate the idea of applying OVP for active systems, we consider a two-dimensional (2D) system 
with a particle confined by four springs of stiffness $k$ and located above a rotary plate in a viscous 
environment, as shown in Fig.~\ref{fig_eg}. 
In this overdamped situation, the mass of the particle can be neglected.
The disk rotates with an angular velocity $\dot{s}=ds/dt$ ($s$ being the angle of rotation) due to a constant 
driving torque $f$ exerted on the disk.
Notice that $f$ can take both positive and negative values and nonzero $f$ breaks the time-reversal symmetry.
When the particle moves from the origin by a displacement $\mathbf{u}=(u_x,u_y)$, the total potential 
energy $U$ and its time derivative can be written as  
\begin{align}
& U =\frac{k}{2}\left(u_x^2+u_y^2 \right)- f s, 
\\
& \dot{U} = k (u_x \dot{u}_x+u_y \dot{u}_y) - f \dot{s}.
\label{confined_partical_energy}
\end{align}

Because the confined particle is located above and close to the rotary disk, linear viscous friction arises 
from the relative motion between the particle and the disk. 
(One can also assume a lubricating layer between the disk and the particle.)
Then the associated dissipation function $\Phi$ is written as
\begin{align}
    \Phi & = \frac{\zeta_u}{2} \left( \dot{u}_x^2 + \dot{u}_y^2 \right)    
    + \frac{\zeta_s}{2} \dot{s}^2 \nonumber \\
    & + \frac{\xi}{2}\left(u_x \dot{s} - \dot{u}_y\right)^2 
    + \frac{\xi}{2}\left(u_y \dot{s} + \dot{u}_x\right)^2,
\label{rotarydisk_dissipation_function}
\end{align}
where $\zeta_u$, $\zeta_s$, and $\xi$ are the positive friction coefficients corresponding to 
the particle velocity, the rotational velocity of the disk, and the relative motion between them, 
respectively.
The last two terms in Eq.~(\ref{rotarydisk_dissipation_function}) arise when the cross product of 
$(0, 0, \dot{s})$ and $(u_x, u_y, 0)$ differs from $(\dot{u}_x, \dot{u}_y, 0)$.
Importantly, these terms indicate that finite $\dot{s}$ induces the quantities $\dot{u}_y/u_x$ and $\dot{u}_x/u_y$ in 
an anti-symmetric way.  
This is the key idea that would later lead to the nonreciprocal interaction of the system, which is 
a common feature of active systems~\cite{YIKLSHK22,Fruchart2021Non-reciprocalTransitions}. 
In general, the two coefficients $\xi$ can be different, but we consider the simplest case here.

The Rayleighian of the system is given by the sum of Eqs.~(\ref{confined_partical_energy}) and (\ref{rotarydisk_dissipation_function}), i.e., $R = \dot{U} + \Phi$~\cite{Doi2011OnsagersMatter,DoiSoftMatterPhysics}.
Minimization of $R$ with respect to $\dot{u}_x$, $\dot{u}_y$, and $\dot{s}$ leads to the 
following set of coupled equations:
\begin{align}
& \zeta_u\dot{u}_x + k u_x + \xi \left(u_y \dot{s} + \dot{u}_x \right) =0, 
\label{eqr1}
\\
&\zeta_u\dot{u}_y + k u_y - \xi \left(u_x \dot{s} - \dot{u}_y \right) =0,  
\label{eqr2}
\\
& \zeta_s \dot{s} + \xi \left( u^2 \dot{s}  -  u_x \dot{u}_y + \dot{u}_x u_y \right) = f, 
\label{eq_s}
\end{align}
where $u^2=u_x^2+u_y^2$ in Eq.~(\ref{eq_s}). 
Equations~(\ref{eqr1}) and (\ref{eqr2}) describe the motion of the particle experiencing both the frictional 
and elastic forces, while the terms with $\xi$ express how the particle velocity couples with the external 
chiral motion $\dot{s}$.
On the other hand, Eq.~(\ref{eq_s}) means that the angular momentum 
$u_x \dot{u}_y - u_y \dot{u}_x$ is generated by the driving torque $f$ through the friction $\xi$.

Next, we use Eq.~(\ref{eq_s}) to eliminate the variable $\dot{s}$ from Eqs.~(\ref{eqr1}) and (\ref{eqr2}).
Then the nonlinear dynamics of $u_x$ and $u_y$ can be expressed in the matrix form as   
\begin{align}
\boldsymbol{\Gamma}
\begin{pmatrix}
	 \dot{u}_x \\ \dot{u}_y
\end{pmatrix} 
= 
\mathbf{E}
\begin{pmatrix}
	u_x \\ u_y
\end{pmatrix}, 
\label{matrix_eq_no_s}
\end{align}
where the friction matrix $\boldsymbol{\Gamma}$ and the generalized elastic matrix $\mathbf{E}$ are introduced by   
\begin{align}
\boldsymbol{\Gamma} =
\begin{pmatrix}
	\zeta_u + \dfrac{ \xi(\xi u_x^2 + \zeta_s)}{\xi u^2 + \zeta_s} & 
	\dfrac{\xi^2 u_x u_y }{\xi u^2+ \zeta_s}
	\\ 	
	\dfrac{\xi^2  u_x u_y }{\xi u^2 + \zeta_s} & 
	\zeta_u + \dfrac{\xi(\xi u_y^2 + \zeta_s)}{\xi u^2 + \zeta_s}
\end{pmatrix}, 
\end{align}
\begin{align}
\mathbf{E} =
\begin{pmatrix}
	- k & - \dfrac{f \xi}{\xi u^2 + \zeta_s}
	\\
	\dfrac{f \xi}{\xi u^2 + \zeta_s} & - k
\end{pmatrix}.
\end{align}
Notice that $\boldsymbol{\Gamma}$ is symmetric and positive-definite while $\mathbf{E}$ is 
anti-symmetric.
Hence, the elimination of $\dot{s}$ gives rise to the nonreciprocal interaction between 
$u_x$ and $u_y$~\cite{YHSK21,YIKLSHK22}. 
We emphasize that the anti-symmetric off-diagonal components of $\mathbf{E}$ are present only 
when $f \neq 0$ corresponding to the nonequilibrium situation.
Moreover, one can easily show that $(u_x,u_y)=(0,0)$ is the unique equilibrium point that is 
globally stable.

For the large-displacement limit, $\xi u^2 \gg \zeta_s$, the above friction and elastic matrices become 
\begin{align}
\boldsymbol{\Gamma} \approx
\begin{pmatrix}
	\zeta_u + \xi u_x^2/u^2 & 
	\xi u_x u_y/u^2
	\\ 	
	\xi  u_x u_y/u^2 & 
	\zeta_u + \xi u_y^2/u^2
\end{pmatrix}, 
\label{largeG}
\end{align}
\begin{align}
\mathbf{E} \approx
\begin{pmatrix}
	- k & - f/u^2
	\\
	 f/u^2 & - k
\end{pmatrix}.
\label{largeE}
\end{align}
The off-diagonal elements of $\mathbf{E}$ become smaller when the displacement amplitude $u^2$ is larger.

On the other hand, for the small-displacement limit, $\xi u^2 \ll \zeta_s$, Eq.~(\ref{matrix_eq_no_s}) can be 
simply linearized as 
\begin{align}
(\zeta_u+\xi) 
\begin{pmatrix}
	\dot{u}_x \\ \dot{u}_y
\end{pmatrix}
= 
\begin{pmatrix}
	-k & -f \xi/\zeta_s
	\\
	f \xi/\zeta_s & -k
\end{pmatrix}
\begin{pmatrix}
	u_x \\ u_y
\end{pmatrix}.
\label{matrix_nos_small_zs}
\end{align}
The frictional factor $\zeta_u+\xi$ on the left-hand side indicates that the energy associated with
$u_x$ and $u_y$ dissipates through the relative motion with respect to the environment 
and also to the rotary disk.
The off-diagonal elements on the right-hand side imply that the degree of nonreciprocality is controlled 
by  the ratio $\xi/\zeta_s$.
This ratio characterizes the relative energy injected to the displacement coordinates 
$u_x$ and $u_y$ with respect to that dissipated through the variable $\dot{s}$.

\begin{figure}[tb]
\centering
\includegraphics[scale=0.7]{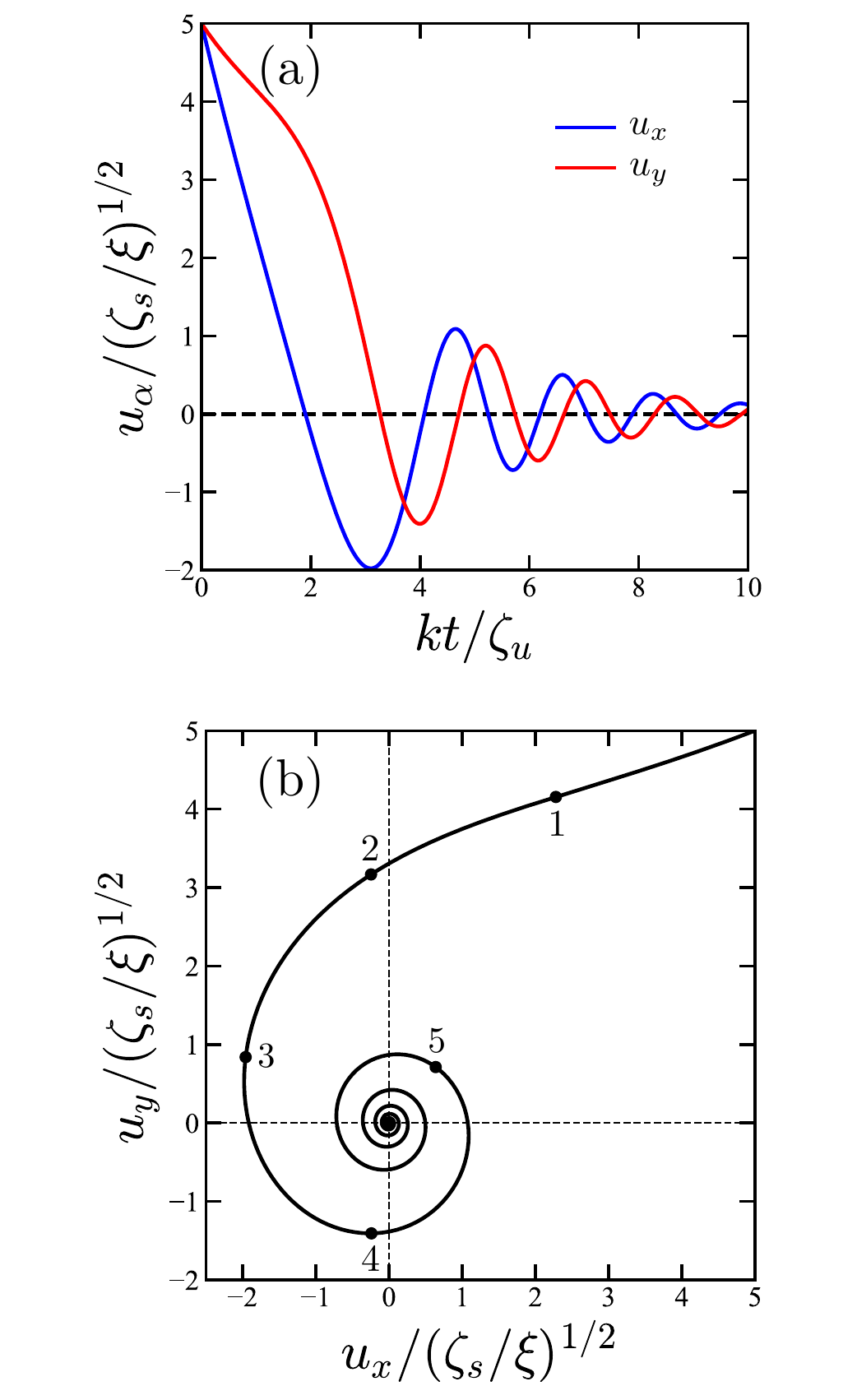}
\caption{(Color online) 
(a) Time evolution of dimensionless $u_x$ (blue) and $u_y$ (red) as a function of dimensionless time
$k t/\zeta_u$ when the initial condition is $u_x /(\zeta_s/\xi)^{1/2}=u_y/(\zeta_s/\xi)^{1/2}=5$ 
[see Eq.~(\ref{matrix_eq_no_s})].
The other parameters are $\xi/\zeta_u= 1.5 $ and $f \xi / (k \zeta_s)=10$.
The oscillatory motion due to the nonreciprocal force starts to appear when 
$u_x$ and $u_y$ become smaller.
(b) The parametric plot of (a) on the plane of $u_x$ and $u_y$.
The numbers at black circles indicate the dimensionless time.}
\label{fig_u1u2}
\end{figure}

The linear equation in Eq.~(\ref{matrix_nos_small_zs}) can easily be solved as 
\begin{align}
u_x = U_0 e^{- \Lambda t} \sin(\Omega t-\phi), \quad 
u_y = U_0 e^{- \Lambda t} \cos(\Omega t-\phi),
\label{u2linearsolution}
\end{align}
where the decay rate $\Lambda$ and the angular frequency $\Omega$ are given by 
$\Lambda = k/(\zeta_u + \xi)$ and $\Omega = f \xi/[(\zeta_u + \xi) \zeta_s]$, respectively.
In the above, the amplitude $U_0$ and the phase difference
$\phi$ are determined by the initial condition. 
The system relaxes to its steady state through the conservative restoring force provided by the 
stiffness $k$ and the dissipation due to the total friction, $\zeta_u + \xi$. 
On the other hand, the oscillatory motion is driven by the torque source $f$ weighted by the 
ratio $\xi/ \zeta_s$.

We choose the units for length and time as $(\zeta_s/\xi)^{1/2}$ and $\zeta_u/k$, respectively.
The other dimensionless parameters are $\xi/\zeta_u$ and $f \xi / (k \zeta_s)=\Omega/\Lambda$.
In Fig.~\ref{fig_u1u2}, we show the time evolution of $u_x$ and $u_y$ by numerically solving 
the nonlinear dynamics in Eq.~(\ref{matrix_eq_no_s}) for certain parameters. 
We see that $u_x$ and $u_y$ do not oscillate when their amplitudes are large, while the 
oscillatory behavior becomes pronounced when the amplitudes become smaller. 
The phase difference between the oscillations of $u_x$ and $u_y$ are almost $\pi/2$.
These behaviors are consistent with the behaviors for the large-amplitude 
[Eqs.~(\ref{largeG}) and (\ref{largeE})] and small-amplitude [Eq.~(\ref{matrix_nos_small_zs})] cases.

Next, we shall extend the above simple model to a 2D continuum model
with odd elasticity. 
Due to the translational symmetry, the free energy of an elastic material should be a functional of the 
gradients of displacement. 
We consider position dependent 2D displacement field $\mathbf{u}(\mathbf{r})=(u_x(\mathbf{r}),u_y(\mathbf{r}))$ 
where $\mathbf{r}=(x,y)$.
When the deformation is small, the free energy $F$ can be expressed in terms of quadratic forms 
of nonsymmetrized strain tensor 
$u_{ij} = \partial u_i/\partial r_j = \partial_j u_i$ ($i, j=x,y$) 
as~\cite{DoiSoftMatterPhysics,LandauElasticity} 
\begin{align}
F =\int d \mathbf{r} \,
\left[\frac{K}{2}w^2+\frac{G}{4}\left( u_{ij} +
u_{ji}-w \delta_{ij} \right)^2-fs\right],
\end{align}
where the summation over repeated indices is assumed, $w = u_{ii}$ is the trace 
of the strain tensor, and $\delta_{ij}$ is the Kronecker delta.
In the above, $K$ and $G$ are the standard bulk and shear moduli, respectively, and $f$ is a constant 
nonequilibrium force conjugate to the extra variable $s(\mathbf{r})$ that is independent of the strain.
For an odd elastic material, $\dot{s}$ must be rotational (with the ability to drive a force field with non-zero curl in the material coordinates), breaking both the time-reversal 
symmetry and the conservative nature of the even elasticity. 
For example, $f$ is the torque density determined by odd microdevices such as internal 
actuators~\cite{Brandenbourger22} and $s$ is the dimensionless microrotation field~\cite{Mitarai2002}.
The time derivative of $F$ can be written as~\cite{DoiSoftMatterPhysics} 
\begin{align}\label{con_energy_change_rate}
\dot{F}=\int{d\mathbf{r}  
\left[\sigma_{ij} \partial_j \dot{u}_i -f\dot{s}\right]},
\end{align}
where $\sigma_{ij} = K w\delta_{ij}+G\left(u_{ij}+u_{ji} - w\delta_{ij}\right)$ 
is the elastic stress tensor.

Next, we discuss the dissipation function of a 2D odd material by using the symmetry argument.
To explain the nonreciprocal relations between strains and strain rates, we introduce the 
following four deformation bases 
$\tau_0 = u_{xx} + u_{yy}$, $\tau_1 = u_{yx} - u_{xy}$, $\tau_2 = u_{yy} - u_{xx}$, 
and $\tau_3 = u_{xy}+u_{yx}$ representing dilation, rotation, and two types of shear, respectively. 
By assuming that the whole system is isotropic, it has been shown that the elastic modulus tensor 
$C_{ijk\ell}$ can be written as~\cite{Scheibner2020OddElasticity,Fruchart22}
\begin{align}
& C_{ijk\ell}  = K\delta_{ij}\delta_{k\ell} - A \epsilon_{ij} \delta_{k\ell} 
- \Xi \delta_{ij} \epsilon_{k\ell} + \Upsilon \epsilon_{ij}\epsilon_{k\ell} 
\nonumber \\
& +  G (\delta_{i\ell} \delta_{jk} + \delta_{ik}\delta_{j\ell} - \delta_{ij}\delta_{k\ell}) 
+ K^\textrm{o} (\epsilon_{ik}\delta_{j\ell} + \epsilon_{j\ell} \delta_{ik}),
\label{isotropic_C}
\end{align}
where $\epsilon_{ij}$ is the 2D Levi-Civita tensor, and $K$ and $G$ are the aforementioned bulk 
and shear moduli, respectively.
For simplicity, we further assume that the solid-body rotations do not induce stress (called objectivity) and set $\Xi=\Upsilon=0$. 
The modulus $A$ converts dilation into torque ($\tau_0 \rightarrow \dot{\tau}_1$), 
and $K^\textrm{o}$ converts shear strains to shear stresses in an antisymmetric way
($\tau_3 \rightarrow \dot{\tau}_2$ and $\tau_2 \rightarrow -\dot{\tau}_3$). 
Both $A$ and $K^\textrm{o}$ correspond to the antisymmetric components of $C_{ijk\ell}$ 
under the exchange of $ij \leftrightarrow k\ell$ and they represent the odd elastic 
moduli~\cite{Scheibner2020OddElasticity,Fruchart22}.

Adapting the above symmetry argument, we consider the dissipation mechanism such that strain 
rates are induced by strains through the rotational field $\dot{s}$.
Notice that $\dot{s}$ conjugates here to the strain because we consider a continuum model with 
internal actuators such as the one shown in Fig.~\ref{fig_eg}.
Among all the possible dissipation terms 
$(\tau_\alpha \dot{s} \pm \dot{\tau}_{\beta})^2$ ($\alpha, \beta=0,\dots, 3$), the symmetry 
properties in Eq.~(\ref{isotropic_C}) (with $\Xi=\Upsilon=0$) lead to the following 
dissipation function
\begin{align}
\Psi  & = \int d\mathbf{r}\, \bigg[ \frac{\zeta_u}{2}  \left( \dot{u}_x^2 + \dot{u}_y^2 \right) 
+\frac{\zeta_s}{2}\dot{s}^2 + \frac{\mu}{2}\left( \tau_0\dot{s} -\dot{\tau}_1 \right)^2
\nonumber \\ 
& + \frac{\nu}{2} \left( \tau_3\dot{s} -\dot{\tau}_2 \right)^2  
+\frac{\nu}{2}\left( \tau_2\dot{s} +\dot{\tau}_3 \right)^2       \bigg].
\label{rotary_con_dissipation_function}
\end{align}
The first two terms are similar to those in Eq.~(\ref{rotarydisk_dissipation_function}).
In the above, we have introduced the positive friction coefficients $\mu$ and $\nu$ that are 
related to the microscopic structure, such as the lattice topology of the odd microdevice~\cite{solidstatephysics}.
The terms that can be obtained by replacing $\dot{s} \rightarrow -\dot{s}$ are not 
included because one of the rotational directions of $\dot{s}$ is specified as in the first model.
We also do not consider the self-driven terms $(\tau_\alpha \dot{s} - \dot{\tau}_{\alpha})^2$ 
since they do not contribute to the odd elasticity.
By adding Eqs.~(\ref{con_energy_change_rate}) and (\ref{rotary_con_dissipation_function}), the Rayleighian 
of the continuum model is now given by $\mathcal{R}=\dot{F} + \Psi$.

Next, we functionally minimize the Rayleighian $\mathcal{R}$ with respect to $\dot{u}_x$, $\dot{u}_y$, and $\dot{s}$, 
and obtain the following set of coupled equations
\begin{align}
& \zeta_u \dot{u}_x  - \partial_i \sigma_{x i}
- \mu \partial_y \left( \tau_0 \dot{s} - \dot{\tau}_1 \right)
\nonumber \\
& - \nu  \Big[ \partial_x \left(\tau_3\dot{s} - \dot{\tau}_2 \right) 
+ \partial_y \left( \tau_2 \dot{s} + \dot{\tau}_3 \right)\Big] = 0, \label{conti_rotary_eq1}
\\
& \zeta_u \dot{u}_y  - \partial_i \sigma_{y i}
+ \mu \partial_x \left( \tau_0 \dot{s} - \dot{\tau}_1 \right)
\nonumber \\
&  + \nu \Big[ \partial_y \left(\tau_3\dot{s} - \dot{\tau}_2 \right) 
- \partial_x \left( \tau_2 \dot{s} + \dot{\tau}_3 \right)\Big] = 0,
\label{conti_rotary_eq2}
\\
&\zeta_s\dot{s} + \mu \left(\tau_0^2\dot{s} - \tau_0\dot{\tau}_1\right)
+ \nu \Big[ \left(\tau_2^2  + \tau_3^2 \right)\dot{s} 
- \dot{\tau}_2\tau_3 + \tau_2\dot{\tau}_3 \Big ]
 = f.
\label{conti_rotary_eq3}
\end{align}
The above equations are nonlinear and cannot be solved analytically. 
To simplify the situation, we focus on the small-strain limit for which the equations can be linearized. 
Then Eq.~(\ref{conti_rotary_eq3}) becomes $\zeta_s \dot{s} \approx f$. 
By substituting this relation into Eqs.~(\ref{conti_rotary_eq1}) and (\ref{conti_rotary_eq2}), we obtain 
the matrix equation of Eq.~(\ref{matrix_eq_no_s}) in which the friction matrix operator 
$\boldsymbol{\Gamma}$ is
\begin{align}
\boldsymbol{\Gamma} =
\begin{pmatrix}
	\zeta_u - \mu \partial^2_y - \nu \nabla^2  & 0
	\\
	0 & \zeta_u - \mu \partial^2_x - \nu \nabla^2 
\end{pmatrix},
\end{align}
where $\nabla^2$ is the 2D Laplacian, and the components of the generalized elastic matrix operator $\mathbf{E}$ are 
\begin{align}
E_{11}&=(K+G) \partial_x^2 + G \partial_y^2 + (f\mu/\zeta_{s})\partial_x\partial_y,
\\
E_{12}&=K \partial_x\partial_y + (f \mu/\zeta_{s}) \partial^2_y + (f \nu/\zeta_{s}) \nabla^2,
\\
E_{21}&=K \partial_x\partial_y - (f \mu/\zeta_{s}) \partial^2_x- (f \nu/\zeta_{s}) \nabla^2,
\\
E_{22}&=(K+G) \partial_y^2 + G \partial_x^2 - (f\mu/\zeta_{s})\partial_x\partial_y.
\end{align}

We now apply the Fourier transform of the obtained linearized equation. 
For an isotropic 2D material, one can choose the $x$-direction as the direction of perturbation without 
loss of generality.
By using the solution of the form $u_x \sim e^{i(qx-\omega t)}$ and 
$u_y \sim e^{-i\omega t}$, where $q$ is wavenumber and $\omega$ is frequency, Eq.~(\ref{matrix_eq_no_s}) can be rewritten in terms of the Fourier components as  
\begin{align}
& i\omega \begin{pmatrix}
	\zeta_u + \nu q^2 &  0
	\\
	0 & \zeta_u + (\mu+\nu) q^2
\end{pmatrix}
\begin{pmatrix}
	u_x \\ u_y
\end{pmatrix} 
\nonumber \\ 
& =
q^2
\begin{pmatrix}
	K+G &  f \nu/\zeta_{s}
	\\
	- f (\mu+\nu)/\zeta_{s} & G
\end{pmatrix}
\begin{pmatrix}
	u_x \\ u_y
\end{pmatrix}. 
\label{conti_rotary_matrix_r1}
\end{align}
The two odd elastic moduli can now be identified as $A= - f \mu/\zeta_{s}$ and 
$K^{\rm o}= f \nu/\zeta_{s}$ (the minus sign of $A$ is corrected as in Ref.~\cite{Fruchart22}), 
which are both proportional to the nonequilibrium force $f$ and the ratio between the friction coefficients (either 
$\mu/\zeta_{s}$ or $\nu/\zeta_{s}$).
Notice that $A$ and $K^{\rm o}$ can take both positive and negative signs depending on the 
sign of $f$.
Hence we have provided a clear physical meaning of the odd elasticity in a nonequilibrium system 
within the OVP framework.
This is the main outcome of our formulation.
Note that the $q^2$-dependence in the friction matrix of Eq.~(\ref{conti_rotary_matrix_r1}) is 
a new contribution.

In the long-wavelength limit where the $q$-dependence on the left-hand side of Eq.~(\ref{conti_rotary_matrix_r1}) 
can be neglected, we obtain the following approximate dispersion relation
\begin{align}
i \omega \approx  
\frac{q^2}{\zeta_u}
\left[ \frac{K}{2} + G \pm \sqrt{\left(\frac{K}{2}\right)^2 
- \frac{f^2 \mu\nu}{\zeta_s^2} -  \left( \frac{f \nu}{\zeta_s} \right)^2} \right].
\label{conti_dispersion}
\end{align}
The instability occurs when the real part of $i\omega$ becomes negative, i.e., when 
\begin{align}
\frac{f \mu}{\zeta_s} <
-\frac{ K G + G^2 + (f \nu/\zeta_s)^2}{f \nu/\zeta_s}.
\end{align}
Moreover, an exceptional point exists when the two eigenvalues of Eq.~(\ref{conti_dispersion}) coincide. 
The oscillatory dynamics takes place when  
$f \mu / \zeta_s >  (K/2)^2 / \left( f \nu/\zeta_s \right) - f \nu/ \zeta_s$~\cite{Scheibner2020OddElasticity}.

To summarize, we have obtained dynamical equations for a nonequilibrium active system with odd 
elasticity within the framework of OVP.
The nonreciprocal equations result from the elimination of the extra variable that is conjugate to the 
nonequilibrium driving force.
We have explicitly shown that the odd elastic moduli are proportional to the nonequilibrium force 
and the friction coefficients.

The concept of the extra nonequilibrium variable and the resulting nonreciprocality can also 
be seen in the studies for micropolar fluid~\cite{Mitarai2002} and odd 
viscosity~\cite{Banerjee2017,Markovich2021}. 
In these works, the extra coordinate represents either the rotating field or the intrinsic rotor.
Different from the approach in Ref.~\cite{Wang2021OnsagersMatter}, we followed the 
framework of OVP to obtain the active dynamics by using the extra nonequilibrium variable. 
This clarifies the energy injection process from the active sources and we have obtained the 
explicit expressions of the odd elastic moduli.

For general nonreciprocal systems, $f$ and $s$ in the free energy do not need 
to stand for specific rotational quantities in the configuration space. 
Any set of generalized forces with their conjugated variables can be chosen as long as they drive 
the system away from the equilibrium and give rise to a rotational trajectory in the state space. 
In an enzymatic system, for example, $f$ and $s$ correspond to the chemical potential 
difference and the chemical reaction variable, respectively~\cite{YIKLSHK22,Yasuda21,Kobayashi22}.
In the future, the microscopic origin of the active quantities such as $f$, $\mu$, and $\nu$ in 
our continuum model needs to be further clarified by using various coarse-graining methods.

\begin{acknowledgements}
We thank M.\ Doi for useful discussion. 
L.S.L.\ is supported by Tokyo Human Resources Fund for City Diplomacy and by Research Institute 
for Mathematical Sciences, an International Joint Usage/Research Center, located in Kyoto University.
K.Y.\ acknowledges the support by a Grant-in-Aid for JSPS Fellows (No.\ 21J00096) from the 
Japan Society for the Promotion of Science (JSPS).
K.I.\ acknowledges the JSPS, KAKENHI for Transformative Research Areas A (No.\ 21H05309) 
and the Japan Science and Technology Agency (JST), PRESTO (No.\ JPMJPR1921).
S.K.\ acknowledges the support by the National Natural Science Foundation of China (Nos.\ 12274098 and 
12250710127) and the startup grant of Wenzhou Institute, University of Chinese Academy of Sciences 
(No.\ WIUCASQD2021041).
\end{acknowledgements}


\end{document}